\documentclass[aps,prb,twocolumn]{revtex4}
\usepackage{makeidx}
\usepackage{amsmath,amssymb,amsfonts,amsthm}
\usepackage{graphicx,bm}

\begin{document}

\title{Hydrogen tunneling in the perovskite ionic conductor BaCe$_{1-x}$Y$%
_{x}$O$_{3-\delta }$ }
\date{}
\author{F. Cordero,$^{1}$ F. Craciun,$^{1}$ F. Deganello,$^2$ V. La Parola,$%
^2$ E. Roncari,$^3$ and A. Sanson$^3$}
\affiliation{$^{1}$ CNR-ISC, Istituto dei Sistemi Complessi, Area della Ricerca di Roma -
Tor Vergata,\\
Via del Fosso del Cavaliere 100, I-00133 Roma, Italy}
\affiliation{$^{2}$ CNR-ISMN, Istituto per lo Studio dei Materiali Nanostrutturati,\\
Via Ugo La Malfa 153, I-90146 Palermo, Italy}
\affiliation{$^{3}$ CNR-ISTEC, Istituto di Scienza e Tecnologia dei Materiali Ceramici,}
\affiliation{Via Granarolo 64, I-48018 Faenza, Italy}

\begin{abstract}
We present low-temperature anelastic and dielectric spectroscopy
measurements on the perovskite ionic conductor BaCe$_{1-x}$Y$_{x}$O$_{3-x/2}$
in the protonated, deuterated and outgassed states. Three main relaxation
processes are ascribed to proton migration, reorientation about an Y dopant
and tunneling around a same O atom. An additional relaxation maximum appears
only in the dielectric spectrum around 60~K, and does not involve H motion,
but may be of electronic origin, \textit{e.g.} small polaron hopping. The
peak at the lowest temperature, assigned to H tunneling, has been fitted
with a relaxation rate presenting crossovers from one-phonon transitions,
nearly independent of temperature, to two-phonon processes, varying as $%
T^{7} $, to Arrhenius-like. Substituting H with D lowers the overall rate by
8 times. The corresponding peak in the dielectric loss has an intensity
nearly 40 times smaller than expected from the classical reorientation of
the electric dipole associated with the OH complex. This fact is discussed
in terms of coherent tunneling states of H in a cubic and orthorhombically
distorted lattice, possibly indicating that only H in the symmetric regions
of twin boundaries exhibit tunneling, and in terms of reduction of the
effective dipole due to lattice polarization.
\end{abstract}

\maketitle


\section{Introduction}

Perovskite cerates and zirconates are a class of materials that, with
appropriate doping, exhibit ionic conductivity both of O vacancies and
protons, and therefore are suitable as solid electrolytes for fuel cells,
gas sensors and other electrochemical devices. Dissolution of H is achieved
in two steps:\cite{Kre97}\ the material is doped with lower valence cations,
\textit{e.g.} BaCe$_{1-x}$Y$_{x}$O$_{3-x/2}$ (BCY) where the partial
substitution of Ce$^{4+}$ with Y$^{3+}$ introduces charge compensating O\
vacancies; the material is then exposed to a humid atmosphere at high
temperature, so that the H$_{2}$O molecules may dissociate, each of them
filling an O vacancy and contributing with two H atoms. Infrared spectroscopy%
\cite{GPB02} and diffraction\cite{Kni00} experiments indicate that H is
bound to an O, forming an OH$^{-}$ ion, but also makes some hydrogen-bonding
with a next-nearest O atom. The proton diffusion is believed to consist of a
rapid rotation about the O\ atom to which is associated, and slower jumps to
one of the eight next-nearest neighbor O atoms with which an instantaneous
hydrogen bond is established (see Fig. \ref{fig7}a below). It has also been
proposed, however, that in some distorted zirconates and titanates the
rotational barrier may be higher than the transfer barrier.\cite{GGJ05} In
various cubic perovskites ABO$_{3}$ there are four equilibrium orientations
of the OH$^{-}$ complex, with an OH separation of $0.9-1.0$\emph{~}\AA\ and
H pointing in the $\left\langle 100\right\rangle $ directions perpendicular
to the B-O-B bond;\cite{MSK96,INI07} in the presence of dopants or non cubic
distortions such positions would be shifted.\cite{Kni00} It is very likely
that the fast local motion of H\ about the same O is dominated by tunneling,
but so far no quantitative measurements of the associated correlation times
have appeared, except for quasi-elastic neutron scattering experiments on
hydrated Ba(Ca$_{0.39}$Nb$_{0.61}$)O$_{2.91}$, where a fast local motion has
been detected above room temperature with an apparent activation energy of $%
\sim 0.1$~eV.\cite{PMS97} Also in SrCe$_{0.95}$Yb$_{0.05}$O$_{2.97}$ a broad
quasielastic component has been attributed to fast proton rotation,\cite%
{MSK96b} but no reliable measurement of the associated rate was possible.

There is some controversy on the effect of dopants on the proton mobility,
as reviewed in Ref. \onlinecite{BSW07}: on the one hand there are several
indications for trapping,\cite{HKM95,HSH98,GLD07, BSW07} with formation of
stable dopant-H complexes, but it has also been proposed that the excess
doped charge distributes over all O sites, causing an increase of the
hopping barrier for the proton over the whole lattice.\cite{MUM04}

Anelastic and dielectric spectroscopies may contribute to answer to such
issues, since both an electric and elastic dipole are associated with a OH$%
^{-}$ ion or Y-OH$^{-}$ complex, and each type of jump or reorientation with
characteristic time $\tau \left( T\right) $ causes a maximum of the losses
at the temperature $T$ and frequency $\omega /2\pi $ such that $\omega \tau
\left( T\right) \simeq 1$. Anelastic relaxation is particularly useful,
since it is almost insensitive to electronic conduction and is not affected
by charged interface layers. We present anelastic and dielectric
spectroscopy measurements on protonated, deuterated and outgassed BaCe$%
_{1-x} $Y$_{x}$O$_{3-x/2}$, where different relaxation processes are
ascribed to H migration, reorientation about an Y dopant and tunneling about
a same O atom; the focus will be on the last type of motion.

\section{Experimental}

The starting powders of BCY were prepared by auto-combustion synthesis,
which is an easy and convenient solution-based method for the preparation of
nanometric mixed-oxide powders. The method\cite{DMD08} is an improvement of
that for metal citrates described in the literature.\cite{XHC04}
Stoichiometric amounts of highly purified metal nitrates were dissolved in
distilled water and mixed with citric acid which acted both as metal ions
complexant and as fuel. The citric acid to metal nitrates ratio was
maintained to 2, ammonium nitrate was added to regulate the fuel to oxidant
ratio (citric acid/total nitrate ions ratio) to 0.4 and ammonia solution
(30\%wt) was added to regulate the pH value at 6. The water solution was
left to evaporate at 80~$^{\mathrm{o}}$C under constant stirring in a beaker
immersed in a heated oil bath, until a whitish and sticky gel was obtained.
The temperature was then raised to 200~$^{\mathrm{o}}$C until the gel became
completely black and dry. The beaker was then put directly on the hot-plate
at $250-300$~$^{\mathrm{o}}$C until the auto-combustion reaction occurred
leaving the powdered product. Crystallization was completed by firing the
combusted powders in stagnant air at 1000~$^{\mathrm{o}}$C for 5~h. No
weight loss occurred during the synthesis and no oxide segregation has been
detected by X-ray diffraction, so that we assume the nominal compositions of
BaCe$_{1-x}$Y$_{x}$O$_{3-x/2}$ with $x=0.1$ (BCY10)\ and $x=0.15$ (BCY15).
The nanopowders were first uniaxially pressed at 50~MPa and then
isostatically pressed at 200~MPa obtaining $60\times 7\times 6$~mm bars,
which were sintered at 1500~$^{\mathrm{o}}$C for 10~h. The bars where cut as
thin reeds about 4~cm long and 1~mm thick, whose major faces where covered
with Ag paint.

The maximum molar concentration $c_{\mathrm{H/D}}$ of H and D was measured
from the change of weight when the sample state was changed between fully
outgassed (up to 730~$^{\mathrm{o}}$C in vacuum $<10^{-5}$~mbar) and
hydrated for 1-2~h at 520~$^{\mathrm{o}}$C in a static atmosphere of $50-100$%
~mbar H$_{2}$O or D$_{2}$O followed by slow cooling; it was found $c_{%
\mathrm{H}}=0.14$ for $x=0.15$ and $c_{\mathrm{H}}\lesssim 0.086$ for $%
x=0.10 $, slightly less than the theoretical maximum $c_{\mathrm{H}}=x$.

The elastic compliance $s\left( \omega ,T\right) =s^{\prime }-is^{\prime
\prime }$ was measured by electrostatically exciting the flexural modes of
the bars suspended in vacuum on thin thermocouple wires in correspondence
with the nodal lines; the 1st, 3rd and 5th modes could be measured, whose
frequencies are in the ratios $1:5.4:13.3$, the fundamental frequencies of
samples being $\omega /2\pi \simeq 2.8$\ kHz. The elastic energy loss
coefficient, or the reciprocal of the mechanical quality factor,\cite{NB72} $%
Q^{-1}\left( \omega ,T\right) =$\ $s^{\prime \prime }/s^{\prime }$ was
measured from the decay of the free oscillations or from the width of the
resonance peak. The elastic compliance $s$\ is the mechanical analogue of
the dielectric susceptibility $\chi $, with $Q^{-1}$\ corresponding to $\tan
\delta $.

The dielectric permittivity $\varepsilon =\varepsilon ^{\prime
}-i\varepsilon ^{\prime \prime }$\ was measured with a HP 4284A impedance
bridge with a four wire probe between 3 and 100~kHz in the same cryostat
used for the anelastic measurements. After depositing the Ag electrodes, an
intense dielectric relaxation process, identified with the motion of charge
carriers within a Schottky barrier at the electrode interface, completely
masks the true bulk relaxation. Such an effect was suppressed by applying
40~V up to $\sim 500$~K, switching the direction of the dc current in order
to avoid electromigration of O vacancies or protons.\cite{Cra08}

A defect hopping or reorienting with characteristic time $\tau $ and causing
a change $\Delta \lambda $ of its elastic quadrupole and $\Delta p$ of its
electric dipole contributes with a Debye peak to the imaginary parts of the
elastic compliance $s$ and dielectric permittivity $\varepsilon $ as:\cite%
{NH65}
\begin{equation}
s^{\prime \prime }=\frac{c~\left( \Delta \lambda \right) ^{2}}{3v_{0}k_{%
\text{B}}T\cosh ^{2}\left( E/2T\right) }\frac{\omega \tau }{1+\left( \omega
\tau \right) ^{2}}\text{ and}  \label{Ds_anel}
\end{equation}

\begin{equation}
\varepsilon ^{\prime \prime }=\frac{c~\left( \Delta p\right) ^{2}}{%
3\varepsilon _{0}v_{0}k_{\text{B}}T\cosh ^{2}\left( E/2T\right) }\frac{%
\omega \tau }{1+\left( \omega \tau \right) ^{2}}~,  \label{De_diel}
\end{equation}%
where $c$ is the molar concentration of relaxing defects, $v_{0}$ the
molecular volume, and $E$ is the energy difference between the states
participating to relaxation. When $E~\gtrsim T$, the higher energy state
becomes less populated and therefore there is a reduced change of the defect
populations under application of the probe field with respect to the case $%
E=0$; the consequent reduction of the relaxation strength is described by
the factor sech$^{2}\left( E/2T\right) $.\cite{33}

\section{Results}

\subsection{Anelastic spectra}

\begin{figure}[tbp]
\includegraphics[width=8.5 cm]{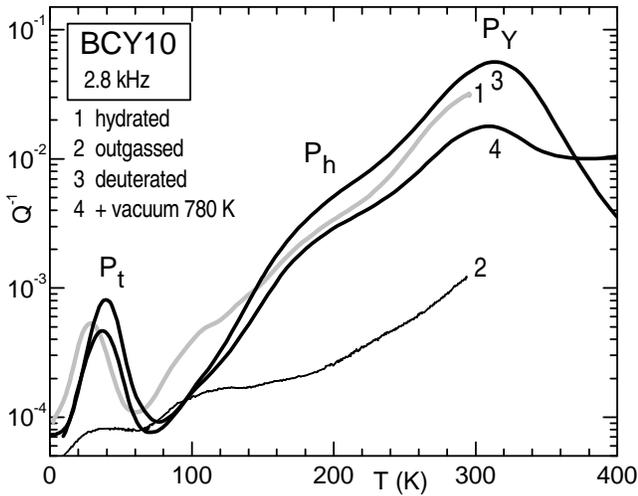}
\caption{Anelastic spectra of BCY10 measured in different conditions of
hydration with H$_{2}$O and D$_{2}$O.}
\label{fig1}
\end{figure}

Figure \ref{fig1} presents the anelastic spectra of BaCe$_{0.9}$Y$_{0.1}$O$%
_{3-\delta }$ measured at the fundamental frequency of 2.8~kHz: 1)\ in the
as prepared state, therefore nearly saturated with H$_{2}$O; 2) after
outgassing H$_{2}$O at 900~$^{\mathrm{o}}$C in a flux of pure O$_{2}$ for
2.5~h; 3)\ after keeping in a static atmosphere of $\sim 70$~mbar D$_{2}$O
at 520~$^{\mathrm{o}}$C followed by cooling at 1~$^{\mathrm{o}}$C/min; 4)\
after having measured in vacuum up to 500~$^{\mathrm{o}}$C with partial loss
of D$_{2}$O.

There are at least three relaxation processes clearly due to the presence of
H or D: the peak labeled P$_{\mathrm{Y}}$ at $\sim 300$~K, P$_{\mathrm{h}}$
at $\sim 200$~K and P$_{\mathrm{t}}$ at $\sim 30$~K; above 400~K start the
contributions due to O\ vacancies. All these peaks shift to higher
temperature when measured at higher frequency, and a preliminary analysis
yields activation energies of 0.58~eV with $\tau _{0}\sim $ $3\times
10^{-14} $~s for P$_{\mathrm{Y}}$ and $\sim 0.4$~eV for P$_{\mathrm{h}}$;
the latter is likely composed of different processes. These peaks are quite
broader than pure Debye relaxations and the possible difference in
relaxation parameters between H and D is not apparent without an accurate
analysis; they must be due to hopping of H between different O\ atoms. Peak P%
$_{\mathrm{t}}$ is the focus of the present work and shifts from 29 to 38~K
when H\ is replaced with D, clearly indicating that it is due to the fast
motion of H about a same O atom with a dynamics dominated by tunneling. An
additional peak at $\sim 100$~K might also be due to H and possibly shifts
to higher temperature after substitution of H with D, but its nature is not
as clear as for the other processes and we will not discuss it further.

\subsection{Comparison between anelastic and dielectric spectra}

\begin{figure}[tbp]
\includegraphics[width=8.5 cm]{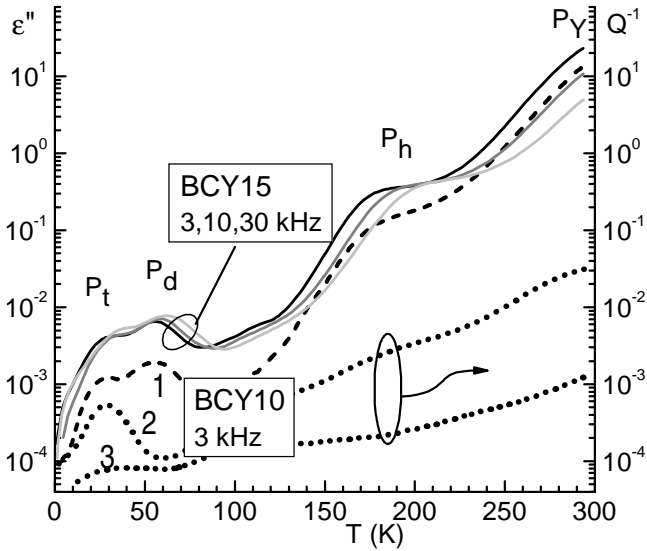}
\caption{Dielectric (left-hand scale) and anelastic (right-hand scale)
spectra of BCY: $\protect\varepsilon ^{\prime \prime }$ of hydrated BCY15 at
3, 10 and 30~kHz (continuous lines) and $\protect\varepsilon ^{\prime \prime
}$ of hydrated BCY10 at 3~kHz (curve 1). The dotted lines are $Q^{-1}$ of
hydrated (2) and outgassed (3) BCY10 measured at 2.8~kHz.}
\label{fig2}
\end{figure}

\begin{figure}[tbp]
\includegraphics[width=8.5 cm]{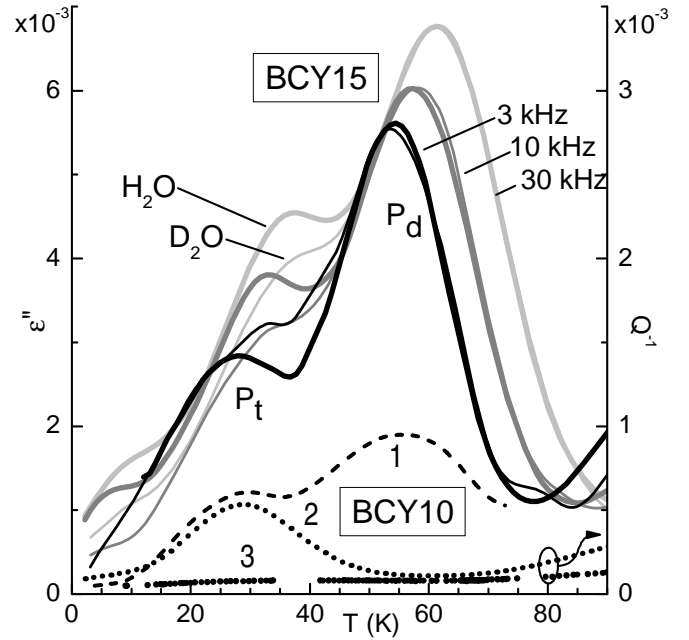}
\caption{Low-temperature region of the dielectric and anelastic spectra of
BCY of Fig. \protect\ref{fig2}. The $\protect\varepsilon ^{\prime \prime }$
(left) and $Q^{-1}$ (right) scales are chosen in order to emphasize the near
coincidence of dielectric and anelastic P$_{\mathrm{t}}$ (curves 1 and 2).}
\label{fig3}
\end{figure}

All the peaks appearing in the anelastic spectrum are present also in the
dielectric one, which also displays an additional maximum P$_{\mathrm{d}}$
at $\sim 60$~K. Figure \ref{fig2} shows $\varepsilon ^{\prime \prime }$ of
hydrated BCY15 and BCY10 ($c_{\mathrm{H}}=0.078$, curve 1); the latter is
compared with the elastic counterpart $Q^{-1}$ in the hydrated (curve 2)\
and outgassed (curve 3) states, measured at the same frequency of 3~kHz. It
appears that the intensities of peaks P$_{\mathrm{t}}$, P$_{\mathrm{h}}$ and
P$_{\mathrm{Y}}$ span more than four orders of magnitude in $\varepsilon
^{\prime \prime }$ but only two in $Q^{-1}$, meaning that the fast tunneling
motion of H produces a much smaller change in the electric dipole than in
the elastic dipole, compared to the hopping motion. On the other hand, the
peak temperatures and activation energies appear to be practically the same
in the dielectric and anelastic losses. These dielectric measurements on
BCY15 have been made after full elimination of the charge relaxation at the
electrodes, so that the intensities of P$_{\mathrm{Y}}$ and P$_{\mathrm{h}}$
are reliable. Figure \ref{fig3} shows $\varepsilon ^{\prime \prime }$ of
both hydrated (thick lines) and deuterated (thin lines)\ BCY15. While peak P$%
_{\mathrm{t}}$ shifts to higher temperature after the isotope substitution
of H with D as for the anelastic case, peak P$_{\mathrm{d}}$ remains
completely unaffected; this fact, together with its absence in the anelastic
losses, indicate that P$_{\mathrm{d}}$ is not connected with the H motion,
but is likely of electronic origin. Curves $1-3$ in Fig. \ref{fig3} are the
same dielectric and anelastic curves of BCY10 appearing in Fig. \ref{fig2},
but this time the two scales are chosen in order to yield the same intensity
of P$_{\mathrm{t}}$. Unfortunately, experimental difficulties connected with
the elimination of the metal-semiconductor barrier at the electrodes did not
allow us to obtain significant dielectric spectra in the outgassed state, so
that it is possible to determine that the intensity of P$_{\mathrm{d}}$
almost triplicates in the hydrated state passing from 10\%\ to 15\%Y doping,
but nothing can be said on its dependence on the content of H or O vacancies.

\section{Discussion}

\subsection{Proton hopping and trapping}

In what follows we will identify peak P$_{\mathrm{Y}}$\ with hopping of H
among the O atoms of YO$_{6}$ octahedra, namely with the reorientation of
the Y-H complex, and P$_{\mathrm{h}}$\ with hopping over CeO$_{6}$
octahedra, with the possible contribution of the formation/dissociation of
Y-H complexes. These assignments are suggested by the fact that it is
natural to assume that (OH)$^{+}$ ions, having a formal charge +1 with
respect to O$^{2-}$, may form relatively stable complexes with trivalent
dopants Y$^{3+}$, having formal charge -1 with respect to Ce$^{4+}$. Then, P$%
_{\mathrm{Y}}$ with higher intensity and activation energy should be due to
the more numerous H atoms associated with Y, while P$_{\mathrm{h}}$ should
be due to the faster jumps of the less numerous H atoms not associated with
Y. The assignment of anelastic relaxation peaks to dopant-H complexes has
already been proposed for BaCe$_{1-x}$Nd$_{x}$O$_{3-\delta }$\cite{Du94} and
(Ba,Sr)Ce$_{1-x}$Yb$_{x}$O$_{3}$,\cite{ZBS95} although the argument that the
(OH)$^{-}$ ion has the same symmetry as the crystal and therefore cannot
produce anelastic relaxation unless forming defect complexes\cite{Du94} is
incorrect. Additional experimental indications that H is trapped by
trivalent dopants in BaCe$_{1-x}$Y$_{x}$O$_{3}$ are the analysis in terms of
two components of the quasielastic neutron scattering peak,\cite{HKM95} and
the observation with EXAFS of an enhancement of the disorder in the
environment of Y after hydration.\cite{GLD07} Also first-principle
calculations and Monte-Carlo simulations of the proton diffusion indicate
that dopants act as traps.\cite{BSW05,BSW07}

On the other hand, there are also experiments suggesting the absence of
significant trapping, like an NMR investigation\cite{MUM04} on BaCe$_{1-x}$Y$%
_{x}$O$_{3}$\ with $x=0.01$\ and 0.1, where the correlation time deduced
from the $^{1}$H NMR relaxation is the same at both doping levels and
reproduces the conductivity with a simple hopping model without trapping. At
this stage, it cannot be completely excluded from our data that there is
indeed very little trapping effect from Y dopants, and the two main
anelastic relaxation processes P$_{\text{Y}}$ and P$_{\text{h}}$ are
associated with H hopping among O1 and O2 atoms of the orthorhombic
structure, having different symmetries. It is possible that H binds to O1
and to O2 with different probabilities, as neutron diffraction experiments
and molecular dynamics simulations suggest,\cite{MKA99,Kni00} and jumps
within the respective sublattices with different rates, so giving rise to
peaks with distinct intensities and temperatures. In the discussion we will
also mention the possibility that in the distorted orthorhombic structure
the reorientation of the OH ions is much slower than found in the higher
temperature phases, and P$_{\text{h}}$\ is due to such slow reorientation.

\subsection{Fit of the anelastic P$_{\mathrm{t}}$}

The relaxation modes of the fast reorientation of OH among four positions
can be found by solving the rate equations for classical hopping between the
four H sites or the quantum mechanical problem with tunneling between
nearest neighbor orientations. In the classical symmetric case there are two
modes: one active in the dielectric relaxation and another in the anelastic
relaxation with a rate twice larger; in the case that the site energies are
different, however, the modes become three, all contributing to both
anelastic and dielectric relaxation, and with possibly widely changing
rates, depending on the type of asymmetry and the degree of coherence of the
eigenstates of H plus polaron-like distortion of the surrounding atoms. At
variance with the geometrically analogous case of the reorientation of Zr-H
complexes in Nb,\cite{36,45,85,CCC98} in highly doped and orthorhombic BCY
it is not possible to distinguish different peaks arising from these modes,
and therefore we will limit ourselves to fit P$_{\mathrm{t}}$ with a single
relaxation time $\tau $ plus broadening, as
\begin{equation}
Q^{-1}\left( \omega ,T\right) =\frac{\Delta _{0}}{T~\text{cosh}^{2}\left(
E/2k_{\text{B}}T\right) }\frac{\sqrt{\alpha \beta }}{\left( \omega \tau
\right) ^{\alpha }+\left( \omega \tau \right) ^{-\beta }},  \label{eq Q0}
\end{equation}%
where the parameters $\alpha ,\beta \leq 1$ produce broadening of the low-
and high-temperature sides of the peak, respectively; when $\alpha =\beta =1$
the above expression reduces to a Debye peak. The parameter $E$ is the
energy difference between the configurations involved in relaxation, and
must be introduced in order to reproduce the enhancement of the peak height
at higher frequency.\cite{33} The relaxation rate was modeled as
\begin{eqnarray}
\tau ^{-1} &=&\tau _{\mathrm{1ph}}^{-1}\coth \left( E/2k_{\text{B}}T\right)
+\left( T/T_{\mathrm{2ph}}\right) ^{n}+  \label{eq t0} \\
&&+\tau _{0}^{-1}\exp \left( -W/T\right) .
\end{eqnarray}%
Such an expression is similar to that used by Kuskowsky, Lim and Nowick,\cite%
{KLN99} for the low temperature dielectric relaxation in Ba$_{1-x}$Nd$_{x}$%
CeO$_{3-x/2}$; it does not rely on a model of the interaction between a
precise defect geometry and the actual phonon bath, but is able to describe
the main hopping regimes, including tunneling in an insulating crystal. One
expects that, starting from low temperature, the transitions between the
defect eigenstates occur through processes involving one-phonon, then
two-phonons and finally several phonons or semiclassical hopping.\cite{CCC98}
In the one-phonon regime the transition rate is $\tau _{\text{1ph}%
}^{-1}\coth \left( E/2k_{\text{B}}T\right) $, which becomes temperature
independent when $k_{\text{B}}T\ $is smaller than the separation $E$ between
the eigenstates; the latter has a form of the type $E\simeq \sqrt{t^{2}+a^{2}%
}$ where the tunneling matrix element $t$ is expected to be smaller than the
typical energy asymmetry $a$ between the site energies due to the
orthorhombic distortion and disordered nature of the BCY solid solution; in
the fit $E=a$. The two-phonon relaxation rate approximately depends on
temperature through a power law,\cite{FS70,Kli83,SBH94} with $5\leq n\leq 9$
depending on the type of interaction with acoustic phonons and on the energy
difference between H\ sites. For interaction with optical phonons of
frequency $\omega _{0}$ an Arrhenius-like $\exp \left( -\hbar \omega _{0}/k_{%
\text{B}}T\right) $ dependence is predicted.\cite{Kli83} At sufficiently
high temperatures an Arrhenius-like temperature dependence is always found,
although the pre-exponential term $\tau _{0}$ is not necessarily related to
the frequency of a local H vibration mode, but depends also on the overlap
of the wave functions of H\ in adjacent sites, and the barrier $W$ may
contain significant corrections due to phonon fluctuations with respect to
the static potential.

\begin{figure}[tbp]
\includegraphics[width=8.5 cm]{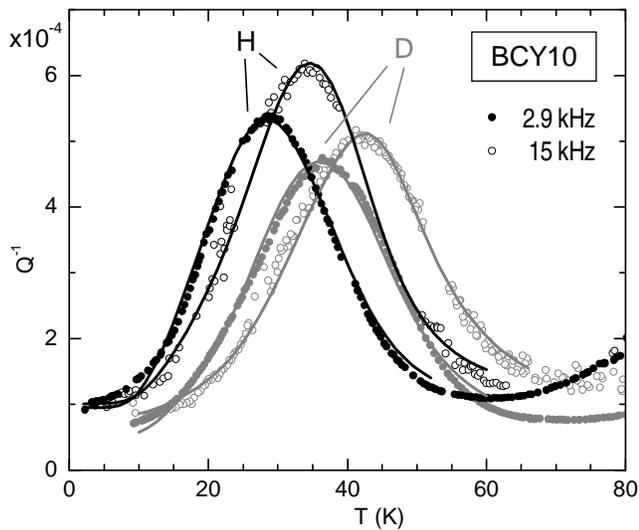}
\caption{Fit of the anelastic P$_{\mathrm{t}}$ of hydrated and deuterated
BCY10, measured at 2.9 and 15~kHz.}
\label{fig4}
\end{figure}

\begin{figure}[tbp]
\includegraphics[width=8.5 cm]{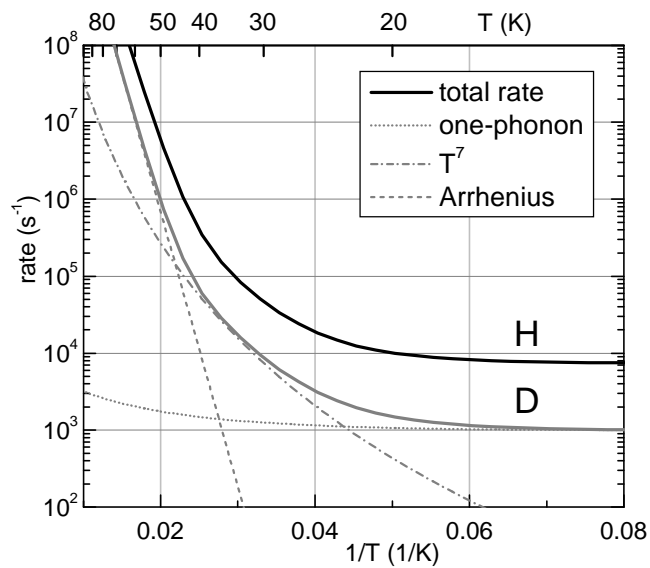}
\caption{Relaxation rates $\protect\tau^{-1}$ used in the fits of Fig.
\protect\ref{fig4}. The rate for deuterium is decomposed into one- ,
two-phonon and Arrhenius-like contributions.}
\label{fig5}
\end{figure}

We found that all the three contributions to $\tau ^{-1}$ are necessary to
obtain a good fit: the one-phonon term reproduces the low temperature
broadening of P$_{\mathrm{t}}$ without resorting to extremely small values
of $\alpha $. The introduction of the power low significantly improves the
fit quality around the maximum. The Arrhenius contribution is not essential
to obtain good fits, since its suppression can be partially compensated by
an increase of $\beta $, namely by diminishing the broadening at high
temperature; however, if one wants to keep $\alpha \simeq \beta $ the
Arrhenius contribution must be included. The condition $\alpha \simeq \beta $
appears desirable, if such parameters should describe the broadening of a
Debye peak due to lattice disorder, and not to other types of interactions,
like collective interactions among H atoms.

The continuous lines in Fig. \ref{fig4} are fits of anelastic P$_{\mathrm{t}%
} $ of BCY10 both hydrated and deuterated with the above expressions plus a
linear background. In view of the partial duplication of the effects of some
parameters, like the high temperature broadening $\beta $ and the Arrhenius $%
\tau _{0}$ and $W$, these fits are not unique, and we tried to obtain a
physically sound combination of the parameters. The mean asymmetry energy is
determined quite precisely as $E/k_{\text{B}}=64$~K from the temperature
dependence of the peak intensities. Regarding broadening, it is possible to
obtain good fits with $\alpha =\beta =0.5$ for both H and D, although $\beta
=0.38$ for H gives a slightly better interpolation, as in Fig. \ref{fig4};
these values of $\alpha $ and $\beta $ imply a broadening that is
conspicuous but expected, in view of the high lattice disorder. The other
parameters are: $\tau _{0}=$ $7.7\times 10^{-14}$~s ($1.1\times 10^{-13}$), $%
W=744$~K (820), $T_{\mathrm{2ph}}=$ $6.7$~K (8.4), $n=7$, $\tau _{\mathrm{1ph%
}}^{-1}=$ 7300~s$^{-1}$ (990) for H (D). The resulting $\tau _{\mathrm{H,D}%
}^{-1}\left( T\right) $ are plotted in Fig. \ref{fig5} and it turns out that
$\tau _{\mathrm{H}}^{-1}/\tau _{\mathrm{D}}^{-1}\simeq 8$ at all
temperatures; such a ratio certainly indicates a non-classical effect of the
isotope mass on the H dynamics. The barrier $W\simeq 800$~K of the
Arrhenius-like contribution is close to the activation energy for rotational
diffusion obtained from quantum molecular dynamics simulations,\cite%
{MKA99,note} and to that extracted from quasi-elastic neutron scattering in
Ba(Ca$_{0.39}$Nb$_{0.61}$)O$_{2.91}$.\cite{PMS97} It cannot be excluded,
however, that it rather originates from two-phonon interaction with optical
phonons,\cite{Kli83} considering that the infrared absorption bands in
various cerates range from $\hbar \omega /k_{\text{B}}=$ 600~K to 1000~K
(400 to 700~cm$^{-1}$).\cite{MOK06} The power law with $n=7$ instead is
typical of two-phonon transitions of asymmetric states interacting with
acoustic phonons; lower values of $n$ yield definitely worse fits, while $%
n\simeq 7.8$ is found if the Arrhenius contribution is omitted.

\subsection{Fit of dielectric P$_{\mathrm{t}}$ and P$_{\mathrm{d}}$}

The most reliable fit is on BCY10, where peak P$_{\mathrm{d}}$ has a reduced
intensity. Below 100~K, $\varepsilon ^{\prime }\left( \omega ,T\right) $ has
already approached the limiting high frequency value $\varepsilon _{\infty
}=20.1$, so that the $\varepsilon ^{\prime \prime }\left( T\right) =$ $%
\varepsilon _{\infty }\times \tan \delta \left( T\right) $ and in Fig. \ref%
{fig6} we show $\tan \delta $ at 10, 30 and 100~kHz; at lower frequency the
noise was too large to add significant information. Note that, due the
imperfect compensation of the cables impedance at low temperature, and the
very small values of the dielectric losses, the $\tan \delta $ curves can be
arbitrarily shifted in the ordinate scale. In fact, it was verified by
switching on and off the cable compensation that below 150~K such a
correction became independent of temperature and therefore introduced only a
shift of the $\tan \delta $ curves.

\begin{figure}[tbp]
\includegraphics[width=8.5 cm]{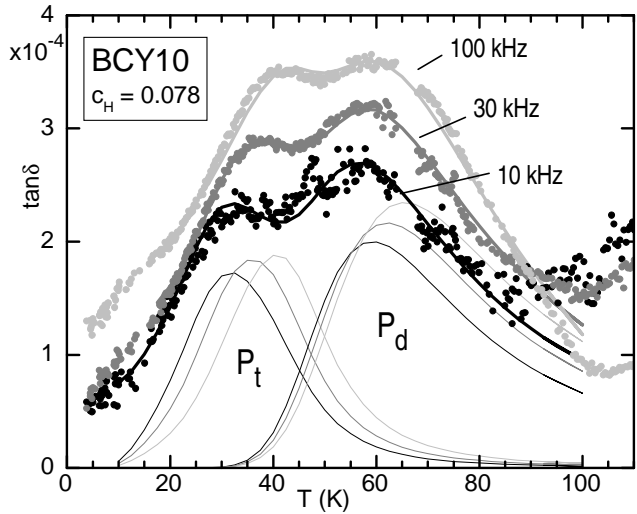}
\caption{Fit of the dielectric spectrum of BCY10.}
\label{fig6}
\end{figure}

The shapes and temperatures of dielectric and anelastic P$_{\mathrm{t}}$
need not to be identical, since the various relaxation modes contributing to
them may have different strengths for anelastic and dielectric relaxation.
Yet, we chose not to let all the parameters of P$_{\mathrm{t}}$ to vary
freely, due to the still overwhelming presence of P$_{\mathrm{d}}$ at higher
temperature and possibly to the presence of another relaxation at lower
temperature. Therefore, all the parameters of P$_{\mathrm{t}}$ where set to
the values of the anelastic fit of Fig. \ref{fig4}, with a factor
multiplying the anelastic $\tau ^{-1}$ as the only degree of freedom: $\tau
_{\mathrm{diel}}^{-1}=r~\tau ^{-1}$; it turns out that $r=10.6$ yields a
good fit. Peak P$_{\mathrm{d}}$, appearing only in the dielectric spectrum,
has an intensity $\Delta \varepsilon _{\mathrm{d}}$ that clearly rises with
temperature (see Fig. \ref{fig3}) and there are two main possible causes for
such a behavior: \textit{i)} P$_{\mathrm{d}}$ is due to the relaxation of a
thermally excited state with energy $E$ over the ground state, \textit{e.g.}
charges from a ionized defect, and therefore the relaxation strength
contains the population of the ionized state as a factor, $\Delta
\varepsilon _{\mathrm{d}}\propto $ $1/\cosh (E/2T)$; \textit{ii)} relaxation
occurs between two states differing in energy by $a$, so that $\Delta
\varepsilon _{\mathrm{d}}\propto $ $n_{1}n_{2}=$ $1/\cosh ^{2}\left(
a/2T\right) $. It is also possible that both mechanisms are present, but the
data do not allow to distinguish between these possibilities, and we will
consider the case $a\neq 0$, $E=0$; note that the two mechanisms give
similar temperature dependence of $\Delta \varepsilon _{\mathrm{d}}$ in a
broad temperature range if $E\sim 1.5a$. The shape of P$_{\mathrm{d}}$ could
be better reproduced with the Cole-Cole expression for broadening, so that
the expression for fitting P$_{\mathrm{d}}$ was
\begin{equation*}
\varepsilon _{\mathrm{d}}^{\prime \prime }=\frac{\Delta _{\mathrm{d}}}{T~%
\text{sech}^{2}\left( a_{\mathrm{d}}/2T\right) }\frac{\sin \left( \pi \alpha
_{\mathrm{d}}/2\right) }{\cosh \left[ \alpha _{\mathrm{d}}\ln \left( \omega
\tau _{\mathrm{d}}\right) \right] +\cos \left( \pi \alpha _{\mathrm{d}%
}/2\right) }
\end{equation*}%
and $\tau _{\mathrm{d}}=\tau _{\mathrm{d0}}\exp \left( W_{\mathrm{d}%
}/T\right) $. The thick curves in Fig. \ref{fig6} are the resulting fit with
linear backgrounds, and also P$_{\mathrm{t}}$ and P$_{\mathrm{d}}$ are
shown. The parameters of P$_{\mathrm{d}}$ are: $a_{\mathrm{d}}=$ 150~K
(150), $\tau _{\mathrm{d0}}=$ $1\times 10^{-15}$~s ($6\times 10^{-15}$), $W_{%
\mathrm{d}}=$ 1270~K (1270) and $\alpha _{\mathrm{d}}=$ 0.26 (0.38), where
the values in parenthesis are obtained from the measurements on BCY15 (data
of Fig. \ref{fig3}). The small values of $\tau _{\mathrm{d0}}$ are in
agreement with the electronic origin of the relaxation, but, considering the
extreme peak broadening (small $\alpha _{\mathrm{d}}$), it cannot be
excluded that a correlated dynamics is present, which may be better
described by a Vogel-Fulcher type $\tau \left( T\right) $ rather than
Arrhenius.

\subsection{The intensity of P$_{\mathrm{t}}$}

\subsubsection{The electric dipole of the OH group}

\begin{figure}[tbp]
\includegraphics[width=8.5 cm]{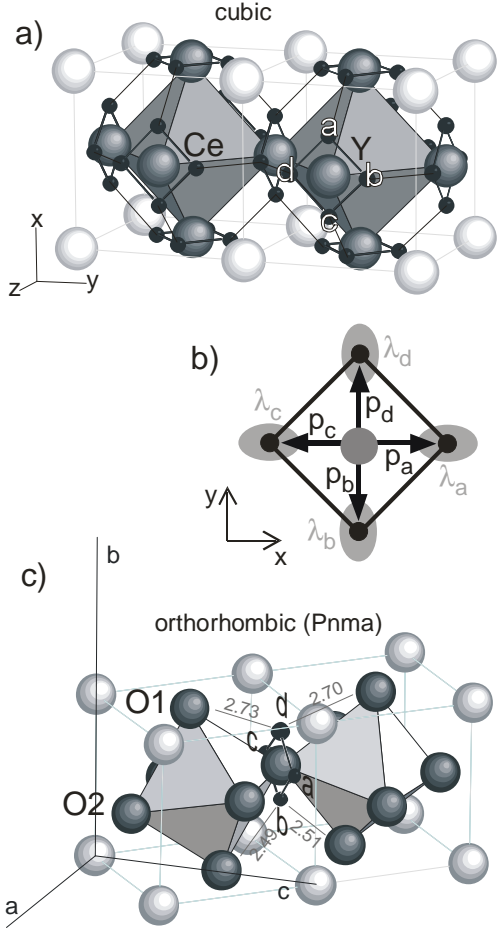}
\caption{(a) Network of the H sites in the cubic structure of BCY; Ba =
white, O = gray, H sites = black, Y/Ce at the centers of the octahedra. (b)
Electric dipoles $\mathbf{p}_{i}^{\mathrm{OH}}$ and elastic quadrupoles $%
\protect\lambda _{i}^{\mathrm{OH}}$ associated with the four orientations of
the OH$^{-}$ ion . (c) Distorted environment of a set of four sites around a
O2 atom in the orthorhombic structure (atomic positions except H from Ref.
\protect\cite{Kni00} in Pnma setting).}
\label{fig7}
\end{figure}

The dependence of the intensity of the anelastic P$_{\mathrm{t}}$ on
the H content and the marked shift to higher temperature with the
heavier D\ isotope mass leave little doubt that anelastic
P$_{\mathrm{t}}$ is due to the fast motion of H around a same O with
a dynamics dominated by tunneling.
It is also clear that P$_{\mathrm{t}}$ has its dielectric counterpart (Fig. %
\ref{fig3}, curves 1 and 2). A puzzling feature of the dielectric P$_{%
\mathrm{t}}$ is its very small intensity, and therefore we discuss now about
the estimated strengths of dielectric and anelastic relaxation. The
dielectric relaxation strength associated with various types of H jumps
should be relatively easy to estimate in a medium which is not particularly
highly polarizable like BCY. Let us first consider P$_{\mathrm{Y}}$,
interpreted as the reorientation of an effective dipole Y$^{\prime }-$OH$%
^{\bullet }$ where the Kr\"{o}ger-Vink notation expresses the fact that Y$%
^{3+}$ has an excess $-e$ charge with respect to Ce$^{4+}$ of the perfect
lattice and OH$^{-}$ has a $+e$ charge with respect to O$^{2-}$. On the time
scale of the H reorientation among different faces of the cube containing Y,
the fast motion around each O is completely averaged out with barycenter
near O, so giving rise to an effective dipole $p^{\mathrm{Y}}=ea/2$, where $%
a $ is the lattice constant of the cubic perovskite (Fig. \ref{fig7}a). A
jump to a different cube face will cause a change of the dielectric dipole
by $\Delta p=e\sqrt{2}a$; setting\cite{Kni00} $a=4.39$~\AA , Eq. (\ref%
{De_diel}) gives a dielectric relaxation strength $\Delta \varepsilon ^{%
\mathrm{Y}}=c_{\mathrm{Y-H}}\times 1060$ at 300~K; therefore the intensity $%
\Delta \varepsilon \simeq 30$ of P$_{\mathrm{Y}}$ at $x=0.15$ (Fig. \ref%
{fig2}) is obtained setting $c_{\mathrm{Y-H}}\simeq 0.03$, namely that at
this temperature $\sim 20\%\ $of H is trapped by Y. It is likely, however,
that $c_{\mathrm{Y-H}}$ is closer to $c_{\mathrm{H}}$ but the effective
dipole $p^{\mathrm{Y}}$ is reduced by the surrounding polarization and
distortion. The dipole involved in P$_{\mathrm{t}}$ instead, is associated
with the OH ion in its four equilibrium positions (Fig. \ref{fig7}b):
\begin{eqnarray}
p_{a}^{\mathrm{OH}} &=&-p_{c}^{\mathrm{OH}}=qd\mathbf{\hat{x},}  \label{pOH}
\\
p_{b}^{\mathrm{OH}} &=&-p_{d}^{\mathrm{OH}}=-qd\mathbf{\hat{y}~,}  \notag
\end{eqnarray}%
where $d\sim 1$~\AA\ is the O-H separation. The actual charge on O and H\ is
likely reduced by polarization effects, and it has been calculated by
atomistic simulations\cite{GIN99} as $-1.426e$ and $0.426e$, so giving $%
q=0.426e$. A\ reorientation of the OH ion by 90$^{\mathrm{o}}$ causes $%
\Delta p^{\mathrm{OH}}=q\sqrt{2}d$ and, setting $E_{a}=64$~K from the fit,
the resulting relaxation strength of peak P$_{\mathrm{t}}$ at $T\simeq 35$~K
should be $\Delta \varepsilon ^{\mathrm{Y}}=c_{\mathrm{Y-H}}\times 41$. It
results that P$_{\mathrm{t}}$ at 35~K should be $\sim 26$ times smaller than
P$_{\mathrm{Y}}$ at 300~K, but instead it is nearly 1000 times smaller,
therefore nearly $40$ times smaller than expected.

\subsubsection{The elastic quadrupole of the OH group}

The anelastic case is different, since the elastic quadrupole
$\lambda $ corresponds to the long range component of the strain
associated with the OH ion and cannot be estimated in an obvious
manner as the electric dipole. It
can only be said that the OH ion in the cubic configuration of Fig. \ref%
{fig7} has orthorhombic symmetry and therefore in a $xy$ cube face it is
\begin{equation}
\lambda _{a}^{\mathrm{OH}}=\lambda _{c}^{\mathrm{OH}}=\left(
\begin{array}{ccc}
\lambda _{1}^{\mathrm{OH}} & 0 & 0 \\
0 & \lambda _{2}^{\mathrm{OH}} & 0 \\
0 & 0 & \lambda _{3}^{\mathrm{OH}}%
\end{array}%
\right)  \label{lOH}
\end{equation}%
and $\lambda _{b,d}^{\text{OH}}$ have the $x$ and $y$ components exchanged.
The reorientation of OH by 90$^{\mathrm{o}}$ causes a change $\Delta \lambda
^{\mathrm{OH}}=\left\Vert \lambda _{1}^{\mathrm{OH}}-\lambda _{2}^{\mathrm{OH%
}}\right\Vert $. On the longer time scale of relaxation P$_{\mathrm{Y}}$,
the four positions of the face $\perp z$ produce a tetragonal quadrupole
\begin{equation}
\lambda ^{z}=\left(
\begin{array}{ccc}
\frac{1}{2}\left( \lambda _{1}^{\mathrm{OH}}+\lambda _{2}^{\mathrm{OH}%
}\right) & 0 & 0 \\
0 & \frac{1}{2}\left( \lambda _{1}^{\mathrm{OH}}+\lambda _{2}^{\mathrm{OH}%
}\right) & 0 \\
0 & 0 & \lambda _{3}^{\mathrm{OH}}%
\end{array}%
\right) ~,
\end{equation}%
plus, in case of Y-OH complex, a possible additional distortion
\begin{equation}
\lambda ^{\mathrm{Y}z}=\left(
\begin{array}{ccc}
\lambda _{2}^{\mathrm{Y}} & 0 & 0 \\
0 & \lambda _{2}^{\mathrm{Y}} & 0 \\
0 & 0 & \lambda _{1}^{\mathrm{Y}}%
\end{array}%
\right) .
\end{equation}%
Then, the anelastic peak P$_{\mathrm{Y}}$ associated with the reorientation
of the tetragonal defect complex Y-OH, has $\Delta \lambda ^{\mathrm{Y}%
}=\left( \lambda _{1}^{\mathrm{Y}}-\lambda _{2}^{\mathrm{Y}}\right) +\lambda
_{3}^{\mathrm{OH}}-\frac{1}{2}\left( \lambda _{1}^{\mathrm{OH}}+\lambda
_{2}^{\mathrm{OH}}\right) $. Lacking any method for estimating these
components of the elastic quadrupoles associated with OH and Y-OH complexes,
we cannot exclude that $\Delta \lambda ^{\mathrm{OH}}$ is $\sim 10$ times
smaller than $\Delta \lambda ^{\mathrm{Y}}$, so determining a reduction of
the intensity of P$_{\mathrm{t}}$ by two orders of magnitude with respect to
P$_{\mathrm{Y}}$. Then, the small intensity of the anelastic P$_{\mathrm{t}}$
does not necessarily constitute a problem as the dielectric intensity does.

\subsubsection{Anelastic and dielectric relaxation strengths of tunneling states in a cubic environment}

The explanation why the intensity of the dielectric P$_{\mathrm{t}}$ is so
much smaller than expected from a simple estimate of the magnitude of dipole
change would be easy if BCY was cubic or very close to cubic. In fact, in
case of coherent tunneling among four nearly equivalent positions, the
effective dipole strength for transitions between H eigenstates would be
much smaller than for hopping among the same positions, at variance with the
anelastic case. The Hamiltonian of the symmetric four-level tunnel system
(FLS) of H may be written in the basis $\left\vert i\right\rangle $ where H
is localized in each site $i=a-d$ as:
\begin{equation}
H_{\mathrm{cub}}=\frac{1}{2}\left[
\begin{array}{cccc}
0 & t & 0 & t \\
t & 0 & t & 0 \\
0 & t & 0 & t \\
t & 0 & t & 0%
\end{array}%
\right] ,  \label{Hcub}
\end{equation}%
where $t$ is the effective tunneling matrix element between adjacent sites
dressed with the interaction with phonons and the site energies are all set
to 0. The eigenstates of $H_{\mathrm{cub}}$ are%
\begin{equation}
\left\vert 1,4\right\rangle \propto \left[
\begin{array}{c}
\pm 1 \\
1 \\
\pm 1 \\
1%
\end{array}%
\right] ,\;\left\vert 2\right\rangle \propto \left[
\begin{array}{c}
0 \\
-1 \\
0 \\
1%
\end{array}%
\right] ,\;\left\vert 3\right\rangle \propto \left[
\begin{array}{c}
-1 \\
0 \\
1 \\
0%
\end{array}%
\right] ~;  \label{Psicub}
\end{equation}%
two states are delocalized over all sites, with energies $E_{1,4}=\mp t/2$,
and two are delocalized over either pair of opposite sites, with energies $%
E_{2,3}=0$. The associated electric dipoles $\mathbf{p}^{\mathrm{FLS}}$ and
elastic quadrupoles $\lambda ^{\mathrm{FLS}}$ may be obtained taking the
matrix elements $\lambda _{i}^{\mathrm{FLS}}=\left\langle i\right\vert
\lambda ^{\mathrm{OH}}\left\vert i\right\rangle $ and $\mathbf{p}_{i}^{%
\mathrm{FLS}}=\left\langle i\right\vert \mathbf{p}^{\mathrm{OH}}\left\vert
i\right\rangle $, with $\lambda ^{\mathrm{OH}}$ and $\mathbf{p}^{\mathrm{OH}%
} $ given above. It is evident that $\mathbf{p}^{\mathrm{OH}}$ averaged over
any of the above states is null, since opposite sites are equally occupied
and cancel out the respective dipoles. Then, no dielectric relaxation is
expected from H tunneling in a cubic environment, or also next to a Y
dopant, which leaves the fourfold symmetry of the FLS. The elastic
quadrupoles, instead, being centrosymmetric as in Eq. (\ref{lOH}), are equal
within pairs of opposite sites, and transitions between the two intermediate
eigenstates cause a change of elastic quadrupole by $\Delta \lambda ^{%
\mathrm{OH}}=\left\Vert \lambda _{1}^{\mathrm{OH}}-\lambda _{2}^{\mathrm{OH}%
}\right\Vert $; transitions between states 2,3 and 1,4 cause a change by $%
\Delta \lambda ^{\mathrm{OH}}/2$ and those between 1 and 4 no anelastic
relaxation. Therefore, the formation of FLS with weak deviations from
fourfold symmetry would explain why the dielectric relaxation from H
tunneling is much more effectively suppressed than the anelastic one, with
respect to classical reorientation.

A picture like this, with H and D performing coherent tunneling within
nearly symmetric FLS,\ has been thoroughly studied by anelastic relaxation
in Nb with substitutional traps.\cite{36,45,85,CCC98}\emph{\ }Also in that
case H tunnels within rings of four equivalent tetrahedral sites on the
faces of the \textit{bcc} cells with a dopant in the center, and performs
overbarrier jumps to the neighboring rings, which form a network exactly as
in the cubic perovskite (see Fig. \ref{fig7}a). In Nb$_{1-x}$Zr$_{x}$ single
crystals with $x=0.0013$ it was also possible to distinguish well separated
relaxation processes arising from transitions involving eigenstates with
different symmetries, only moderately perturbed by interactions among
dopants;\cite{85} in addition, the dependence of the anelastic relaxation
due to the slower reorientation among different cube faces on the symmetry
of the excitation stress\cite{45} provides evidence that the symmetry of the
FLS persists at least up to 150~K, implying that the H eigenstates maintain
coherence up to that temperature. It should be noted that the measured
effective tunneling matrix element between tetrahedral sites in Nb is $t\sim
0.2$~meV for H and 0.02~meV for D, corresponding to $t/k_{\text{B}}=2$ and
0.2~K or $t/h=4\times 10^{10}$ and $4\times 10^{9}$~s$^{-1}$ respectively;
it appears therefore that coherence is maintained both at temperatures
orders of magnitude larger than $t/k_{\text{B}}$ and with tunneling
frequencies of the order of phonon frequencies.

A theoretical basis for the persistence of quantum coherence at such high
temperatures comes from analytical and numerical analysis of the
centrosymmetric FLS, where coherent oscillations of the H\ populations are
found even for strong interaction with the thermal bath,\cite{WG00} and from
the analysis of the dynamics of polarons using the dynamical mean-field
approximation,\cite{FC03} which is a non-perturbative approach. It is shown
that the coherence of the state including tunneling particle and surrounding
polaronic distortion is maintained to temperatures up to a substantial
fraction of the energy of the phonon coupled with the particle, $k_{\text{B}%
}T\sim 0.2\hbar \omega _{0}$, also in the case of strong coupling, where the
energy for the polaron formation is $E_{p}>\hbar \omega _{0}$. We assume $%
E_{p}\simeq 1$~eV from the theoretical estimate\cite{SBW07} of the
self-trapping energy of H in BaZrO$_{3}$, and that the strongest coupling is
with the O-Ce-O bending mode (in BaCeO$_{3}$ $\hbar \omega _{0}=$ 41~meV\cite%
{CPL99}) modulating the distance with the neighboring O atoms and therefore
the hydrogen bonds with them. Then we are in the limit of strong coupling,
and the dynamical mean-field analysis\cite{FC03} ensures us that coherent
states may be maintained up to $0.2\hbar \omega _{0}/k_{\text{B}}\simeq $
100~K; this holds for both H\ and D, since\ the smallness of the dressed
tunneling matrix element should not be a problem, until the tunneling
frequency is larger than the measuring frequency.

\subsubsection{Tunneling states in the orthorhombic lattice}

The main problem with the above explanation of the smallness of the
dielectric relaxation strength of P$_{\mathrm{t}}$ is that BCY at low
temperature is not cubic but orthorhombic, and the FLS should be far from
symmetric. Figure \ref{fig7}c shows an undistorted ring, as in the cubic
case, between two octahedra tilted as in the orthorhombic $Pnma$ low
temperature structure of BaCe$_{0.9}$Y$_{0.1}$O$_{3-\delta }$.\cite{Kni00}
The O\ atoms split in two types: O1 at the vertices of the octahedra along
the $b$ axis and O2 near the $ac$ plane; neutron diffraction indicates that,
at 4.2~K, H in BCY occupies a site near the one labeled as $d$ in Fig. \ref%
{fig7}c, which is also the one with the largest distances from the next
nearest neighbor O\ atoms. It is therefore reasonable to assume that in the
distorted orthorhombic structure H occupies sites slightly displaced from
those in the cubic cell, and that the site energy is mainly determined by
the mean distance $\overline{l}$ from the two next nearest neighbor O atoms,
with which some hydrogen bonding can take place.\cite{BSW07} For the sites
labeled $a-d$ in Fig. \ref{fig7}b it is $\overline{l}=$ 2.66, 2.71, 2.53,
2.50 respectively, so that the site energies may be written as $E_{a}\simeq
E_{b}=a/2$, $E_{c}\simeq E_{d}=-a/2$, and the Hamiltonian of the tunnel
system becomes:%
\begin{equation}
H_{\mathrm{ortho}}=\frac{1}{2}\left[
\begin{array}{cccc}
a & t & 0 & t \\
t & a & t & 0 \\
0 & t & -a & t \\
t & 0 & t & -a%
\end{array}%
\right] ,
\end{equation}%
with eigenstates%
\begin{equation}
\left\vert 1,2\right\rangle \propto \left[
\begin{array}{c}
a-\delta \\
\mp \left( a-\delta \right) \\
\mp t \\
t%
\end{array}%
\right] ,\;\left\vert 3,4\right\rangle \propto \left[
\begin{array}{c}
a+\delta \\
\mp \left( a+\delta \right) \\
\mp t \\
t%
\end{array}%
\right] ,
\end{equation}%
and energies $E_{1,2}=\left( -\delta \mp t\right) /2,$ $E_{3,4}=\left(
\delta \mp t\right) /2,$ where $\delta =\sqrt{a^{2}+t^{2}}$. This means
that, when $a\gg t$, there are two low energy eigenstates with H\ mainly
delocalized over sites $c$ and $d$ and two higher energy eigenstates
delocalized over sites $a$ and $b$; these states have an electric dipole $%
p\simeq ed/\sqrt{2}$ oriented roughly midway between the two occupied sites.
Therefore, while in the limit $a\ll t$ , valid for the cubic ideal case, the
H atom is delocalized over all sites and the electric dipole is averaged out
to almost zero, H in BCY at low temperature is expected to be in the
opposite limit $a\gg t$ where the averaging effect of the electric dipole
occurs only within the pairs $ab$ and $cd$ of low- and high-energy states,
without a suppression of the dipole magnitude.

The present data do not allow an estimate of $t$, neither do we know of
estimates of the energy asymmetry $a$ due to the orthorhombic distortion,
but based on the comparison with the better known case of Zr-H complexes in
Nb we should be in the limit $a\gg t$. In fact, $t$ should be smaller than
in Nb, since, assuming an O-H bond $\simeq 0.93$~\AA\ long, the distance
between neighboring sites is $\sim 1.3$~\AA\ in BCY while it is only $1$~%
\AA\ in Nb; in addition, the maximum of the anelastic relaxation is shifted
to higher temperature with respect to Nb$_{1-x}$Zr$_{x}$H$_{y}$, indicating
slower transition rates; therefore it should be $t/k_{\text{B}}\ll 1$~K, the
value found in Nb. On the other hand, we expect $a/k_{\text{B}}\gtrsim
10^{2} $~K, considering that the random strains due to $<1$ at\% of
impurities in Nb cause $a/k_{\text{B}}$ of tens of kelvin,\cite{CCC98} and
the strain associated with the octahedral tilts in orthorhombic BCY is
certainly larger. Another indication that $a$ due to the orthorhombic
distortion is large comes from the theoretical estimate\cite{SBW07} $%
E_{p}\simeq 1$~eV of the self-trapping energy of H in BaZrO$_{3}$.

\subsubsection{Symmetric tunneling states within twin walls}

After these arguments, it is puzzling that the intensity of the dielectric P$%
_{\mathrm{d}}$ in orthorhombic BCY is so small. The situation is similar to
that found in hydrated BaCe$_{1-x}$Nd$_{x}$O$_{3-\delta }$, where a low
temperature dielectric relaxation exists whose rate has a temperature
dependence indicating tunneling; however, it was concluded that if all the
protons were responsible for such a relaxation, the intensity should be 50
times larger, and therefore only special defect configurations contributed
to that relaxation.\cite{KLN99} In the present case, unless the OH dipole is
smaller than estimated,\cite{GIN99} less than 3\%\ of the H atoms should
contribute to P$_{\mathrm{t}}$ and it should be explained what kind of
particular configuration exhibits tunneling and what is the dynamics of the
majority H atoms. In fact, both neutron spectroscopy at high temperature\cite%
{MSK96b,PMS97} and simulations\cite{BSW07} indicate that the reorientation
of the OH ion is much faster than the hopping between different O atoms in
perovskite cerates. A possible scenario is that the fast reorientation
occurs only in high temperature phases that are cubic or less distorted than
the orthorhombic phase, whereas in the latter H is nearly localized at
lowest energy site, close to site $d$ in Fig. \ref{fig7}b. Then the
reorientation of the OH ion would be slower than that producing P$_{\mathrm{t%
}}$ and might be identified with P$_{\mathrm{h}}$ or some broad peak masked
by the tail of P$_{\mathrm{Y}}$ and by P$_{\mathrm{h}}$, perhaps the one
around 100~K in Fig. \ref{fig1}. In this scenario the tunneling motion would
appear only in particularly symmetric environments, \textit{e.g.} at the
boundaries between different structural domains. We are not aware of any
study of the density, width and morphology of the domain boundaries in BCY,
but indirect support to this mechanism comes from a simulation on
orthorhombic CaTiO$_{3}$, where the twin walls are found to be about 6
pseudocubic cells wide and to trap the O vacancies;\cite{CDS03} it is also
proposed that the diffusion of O vacancies should be faster within the twin
planes, which are more symmetric than the orthorhombic bulk. Also in the
case of H in BCY, the reorientation rate of H might be faster at the twin
walls, but the relevance of this effect to the long range mobility would be
limited, since the rate limiting step is not the OH reorientation but the
hopping to a different O\ atom. We think however that the effect of the
greater symmetry at the twin walls should be studied also in relation with
the hopping mechanism and in the high temperature phases of BCY. Also
simulations on BaZrO$_{3}$\ and CaTiO$_{3}$\ suggest that the octahedral
tilting is essential in determining H site energies and diffusion paths.\cite%
{GGJ05}

If H tunneling in the orthorhombic phase indeed occurs only in the twin
boundaries, than the intensity of peak P$_{\mathrm{t}}$ would depend on
their density, which in turn may depend on microstructure and thermal
history. This might be the reason why the dielectric P$_{\mathrm{t}}$ in
BCY15 is nearly three times more intense than in BCY10, a feature otherwise
difficult to explain; additional measurements are necessary to clarify this
issue.

Another factor that may contribute to lower the dielectric relaxation
strength of P$_{\mathrm{t}}$ is the lattice polarization around the OH$^{-}$
ion; in fact, the actual dipole is not the bare OH$^{-}$ dipole with nominal
charges $\pm 1$, but it is due to the OH$^{-}$ complex plus the shifted
surrounding atoms and with charge transfers with respect to the purely ionic
case. We estimated the intensity of dielectric P$_{\mathrm{t}}$ assuming the
OH dipole calculated in Ref. \cite{GIN99}, but the actual dipole may be even
smaller.

\section{ Conclusions}

Three main relaxation processes have been identified both in the anelastic
and dielectric spectra as due to hopping of H around an Y dopant (peak P$_{%
\mathrm{Y}}$), hopping far from dopants (peak P$_{\mathrm{h}}$) and
tunneling within the four sites around a same O atom (peak P$_{\mathrm{t}}$%
). An additional dielectric relaxation maximum around 60~K (P$_{\mathrm{d}}$%
) does not involve H motion, but rather appears as a relaxation of
electronic origin like small polaron hopping. Peak P$_{\mathrm{t}}$ can be
fitted assuming a relaxation rate that is Arrhenius-like above $\sim 50$~K,
possibly due to two-phonon transitions with optical phonons rather than to
overbarrier hopping, below 50~K exhibits a $T^{7}$ dependence typical of
two-phonon transitions with acoustic phonons, and finally becomes nearly
constant below $\sim 20$~K, as expected from one-phonon transitions. The
isotopic substitution with D decreases the rate by a factor of $8$. The
dielectric spectra are more difficult to analyze, due to the presence of
peak P$_{\mathrm{d}}$ and possibly other peaks at lower temperature, but
reasonable fits are obtained using the same anelastic expression of P$_{%
\mathrm{t}}$ with a rate increased by $\sim 10$ times; such a difference in
the rate may be due to the fact that, although not clearly distinguishable,
there are at least three anelastic and dielectric relaxation modes
contributing to P$_{\mathrm{t}}$, having different strengths and rates.

The intensity of the dielectric P$_{\mathrm{t}}$ is nearly 40 times smaller
than expected from a simple estimate of the reorientation of the electric
dipole associated with the OH$^{-}$ ion. It is shown that, while this
suppression of the dielectric relaxation would be easily explained in terms
of coherent tunneling of H around O in a cubic environment, the same seems
not to be true in the presence of the low-temperature orthorhombic
distortion. As alternative or concomitant explanations, it is proposed that
H\ tunneling may occur only in the more symmetric cells within twin walls,
while slower semiclassical reorientation of OH$^{-}$ would occur within the
orthorhombic domains; in addition, the total dipole of OH$^{-}$ ion and
surrounding lattice polarization may be smaller than expected.

\subsection*{Acknowledgments}

We wish to thank S. Ciuchi for useful discussions on the coherence of the
polaron states, O. Frasciello for valuable suggestions on the dielectric
measurements, and F. Corvasce, M. Latino, A. Morbidini for their technical
assistance. This research is supported by the FISR Project of Italian MIUR:
"Celle a combustibile ad elettroliti polimerici e ceramici: dimostrazione di
sistemi e sviluppo di nuovi materiali".


\begin{references}
\bibitem{Kre97} K.D. Kreuer, Solid State Ion. \textbf{97}, 1 (1997).

\bibitem{GPB02} M. Glerup, F.W. Poulsen and R.W. Berg, Solid State Ion.
\textbf{148}, 83 (2002).

\bibitem{Kni00} K.S. Knight, Solid State Ion. \textbf{127}, 43 (2000).

\bibitem{GGJ05} M.A. Gomez, M.A. Griffin, S. Jindal, K.D. Rule and V.R.
Cooper, J. Chem. Phys. \textbf{123}, 94703 (2005).

\bibitem{MSK96} W. M{\"u}nch, G. Seifert, K.D. Kreuer and J. Maier, Solid
State Ion. \textbf{86- 88}, 647 (1996).

\bibitem{INI07} T. Ito, T. Nagasaki, K. Iwasaki, M. Yoshino, T. Matsui, H.
Fukazawa, N. Igawa and Y. Ishii, Solid State Ion. \textbf{178}, 607 (2007).

\bibitem{PMS97} M. Pionke, T. Mono, W. Schweika, T. Springer and H. Schober,
Solid State Ion. \textbf{97}, 497 (1997).

\bibitem{MSK96b} Th. Matzke, U. Stimming, Ch. Karmonik, M. Soetratmo, R.
Hempelmann and F. G{\"u}thoff, Solid State Ion. \textbf{86}, 621 (1996).

\bibitem{BSW07} M.E. Bj{\"o}rketun, P.G. Sundell and G. Wahnstr{\"o}m, Phys. Rev. B
\textbf{76}, 054307 (2007).

\bibitem{HKM95} R. Hempelmann, Ch. Karmonik, Th. Matzke, M. Cappadonia, U.
Stimming, T. Springer and M.A. Adams, Solid State Ion. \textbf{77}, 152
(1995).

\bibitem{HSH98} R. Hempelmann, M. Soetratmo, O. Hartmann and R. W{\"a}%
ppling, Solid State Ion. \textbf{107}, 269 (1998).

\bibitem{GLD07} F. Giannici, A. Longo, F. Deganello, A. Balerna, A.S. Arico
and A. Martorana, Solid State Ion. \textbf{178}, 587 (2007).

\bibitem{MUM04} H. Maekawa, Y. Ukei, K. Morota, N. Kashii, J. Kawamura and
T. Yamamura, Solid State Commun. \textbf{130}, 73 (2004).

\bibitem{DMD08} F. Deganello, G. Marc{\`{\i}} and G. Deganello, submitted to
J. Eur. Ceram. Soc.

\bibitem{XHC04} Q. Xu, D.-P. Huang, W. Chen, J.-H. Lee, H. Wang and R.-Z.
Yuan, Scripta Mater. \textbf{50}, 165 (2004).

\bibitem{NB72} A.S. Nowick and B.S. Berry, \textit{Anelastic Relaxation in
Crystalline Solids}. (Academic Press, New York, 1972).

\bibitem{Cra08} F. Craciun, unpublished.

\bibitem{NH65} A.S. Nowick and W.R. Heller, Adv. Phys. \textbf{14}, 101
(1965).

\bibitem{33} F. Cordero, Phys. Rev. B \textbf{47}, 7674 (1993).

\bibitem{Du94} Y. Du, J. Phys. Chem. Sol. \textbf{55}, 1485 (1994).

\bibitem{ZBS95} L. Zimmermann, H.G. Bohn, W. Schilling and E. Syskakis,
Solid State Ion. \textbf{77}, 163 (1995).

\bibitem{BSW05} M.E. Bj{\"o}rketun, P.G. Sundell, G. Wahnstr{\"o}m and D.
Engberg, Solid State Ion. \textbf{176}, 3035 (2005).

\bibitem{MKA99} W. M{\"u}nch, K.D. Kreuer, ST. Adams, G. Seifert and J.
Maier, Phase Transitions \textbf{68}, 576 (1999).

\bibitem{36} G. Cannelli, R. Cantelli, F. Cordero and F. Trequattrini, Phys.
Rev. B \textbf{49}, 15040 (1994).

\bibitem{45} G. Cannelli, R. Cantelli, F. Cordero, F. Trequattrini and H.
Schultz, J. Alloys and Compounds \textbf{231}, 274 (1995).

\bibitem{85} F. Cordero, A. Paolone and R. Cantelli, J. Alloys and Compounds
\textbf{330}, 467 (2002).

\bibitem{CCC98} \textit{Tunneling Systems in Amorphous and Crystalline Solids%
}. ed. by P. Esquinazi (Springer, Berlin, 1998).

\bibitem{KLN99} I. Kuskovsky, B.S. Lim and A.S. Nowick, Phys. Rev. B \textbf{%
60}, R3713 (1999).

\bibitem{FS70} C.P. Flynn and A.M. Stoneham, Phys. Rev. B \textbf{1}, 3966
(1970).

\bibitem{Kli83} M.I. Klinger, Phys. Rep. \textbf{94}, 183 (1983).

\bibitem{SBH94} V. Storchak, J.H. Brewer, W.N. Hardy, S.R. Kreitzman and
G.D. Morris, Phys. Rev. Lett. \textbf{72}, 3056 (1994).

\bibitem{note} In Ref. \onlinecite{MKA99} the barrier is calculated assuming that
the proton is a classical particle. Yet, the comparison with $W$ in Eq. (\ref{eq t0})
should be meaningful, since $W$ is the effective activation energy of the fast
H motion in the high temperature limit, which eventually becomes overbarrier hopping.

\bibitem{MOK06} A. Mineshige, S. Okada, M. Kobune and T. Yazawa, Solid State
Ion. \textbf{177}, 2443 (2006).

\bibitem{GIN99} R. Gl{\"o}ckner, M.S. Islam and T. Norby, Solid State
Ion. \textbf{122}, 145 (1999).

\bibitem{WG00} M. Winterstetter and M. Grifoni, Phys. Rev. B \textbf{62},
3237 (2000).

\bibitem{FC03} S. Fratini and S. Ciuchi, Phys. Rev. Lett. \textbf{91},
256403 (2003).

\bibitem{SBW07} P.G. Sundell, M.E. Bj{\"o}rketun and G. Wahnstr{\"o}m, Phys.
Rev. B \textbf{76}, 094301 (2007).

\bibitem{CPL99} I. Charrier-Cougoulic, T. Pagnier and G. Lucazeau, J. Solid
State Chem. \textbf{142}, 220 (1999).

\bibitem{CDS03} M. Calleja, M.T. Dove and E.H. Salje, J. Phys.: Condens.
Matter \textbf{15}, 2301 (2003).
\end{references}

\end{document}